\documentclass[%
 aip,
 jmp,%
 amsmath,amssymb,
 reprint,%
]{revtex4-1}
\usepackage{subfigure}

\usepackage{graphicx}
\usepackage{dcolumn}
\usepackage{bm}

\begin{document}
\title[BPP profile measurements in MAST]{Profile measurements in the plasma edge of MAST using a ball pen probe}
\author{ N R Walkden$^{1,2}$, J Adamek$^{3}$, S Allan$^{1}$, B D Dudson$^{2}$, S Elmore$^{1}$, G Fishpool$^{1}$, J Harrison$^{1}$, A Kirk$^{1}$, M Komm$^{3}$ and the MAST team$^{1}$
        \\ \small{$^{1}$ EURATOM/CCFE Fusion Association, Culham Science Centre, Abingdon, OX14 3DB, UK} 
        \\ \small{$^{2}$ York Plasma Institute, Department of Physics, University of York, Heslington, York, YO10 5DD, UK} 
        \\ \small{$^{3}$ Institute of Plasma Physics of AS CR, v. v. i., Za Slovankou 3, 182 00 Praha 8, Czech Republic}
        \\ Email: \texttt{nrw504@york.ac.uk} }
\date{}

\begin{abstract}
The ball pen probe (BPP) technique is used successfully to make profile measurements of plasma potential, electron temperature and radial electric field on the Mega Amp Spherical Tokamak (MAST). The potential profile measured by the BPP is shown to significantly differ from the floating potential both in polarity and profile shape. By combining the BPP potential and the floating potential the electron temperature can be measured, which is compared with the Thomson scattering (TS) diagnostic. Excellent agreement between the two diagnostics is obtained when secondary electron emission is accounted for in the floating potential. From the BPP profile an estimate of the radial electric field is extracted which is shown to be of the order $\sim$ 1kV/m and increases with plasma current. Corrections to the BPP measurement, constrained by the TS comparison, introduce uncertainty into the $E_{R}$ measurements. The uncertainty is most significant in the electric field well inside the separatrix. The electric field is used to estimate toroidal and poloidal rotation velocities from $\textbf{E}\times\textbf{B}$ motion. This paper further demonstrates the ability of the ball pen probe to make valuable and important measurements in the boundary plasma of a tokamak.  
\end{abstract}

\pacs{...}
\maketitle

\section{Introduction}
The plasma potential plays an important role in many of the processes occurring in the tokamak plasma edge. In the drift ordering of the two-fluid equations appropriate to the plasma edge \cite{Braginskii,SimakovPoP2003,SimakovPoP2004} the $\textbf{E}\times\textbf{B}$ velocity, with $\textbf{E} = -\nabla\phi$ where $\phi$ is the plasma potential,  provides the dominant advective flow and determines the dynamics of both turbulence \cite{MilitelloPPCF2013} and individual filaments \cite{WalkdenPPCF2013}. On the equilibrium scale gradients in the radial electric field can lead to a strongly sheared flow close to the separatrix, commonly termed the radial shear layer. Despite its important role in edge physics the plasma potential has remained a difficult quantity to measure accurately on either the fluctuation or equilibrium scale. The ball pen probe (BPP), developed by Adamek \emph{et.al} \cite{AdamekCJP2004,AdamekCJP2005} offers a diagnostic technique capable of measuring the plasma potential with a robust implementation that may withstand particle and heat fluxes from the plasma at radii up to and within the separatrix. The BPP technique has previously been tested simultaneously with an emissive probe \cite{AdamekCJP2005} and a self-emitting Langmuir probe \cite{AdamekPreprint} and shows excellent agreement in each case. The robustness of the BPP design against the high particle and heat fluxes in the plasma edge, as well as its relative simplicity in design makes it an attractive option over the emissive probe, which can rarely be operated near to the separatrix due to its inherently weak structure. The BPP technique has been used on the CASTOR \cite{AdamekCJP2004,AdamekCJP2005,SchritweisserCJP2006}, ASDEX-Upgrade \cite{AdamekJNM2009,AdamekCPP2010,HoracekNF2010}, ISTTOK \cite{SilvaEPS2013} and COMPASS \cite{AdamekEPS2014,AdamekCPP2014} tokamaks and in the torsatron TJ-K, alongside two other low temperature plasma devices \cite{AdamekCPP2013}. 
\\In this paper results are presented from the development and operation of a BPP on the Mega Amp Spherical Tokamak (MAST)\cite{MAST}. MAST, being a tight aspect ratio tokamak, has a reduced magnetic field strength on the low field side compared to both ASDEX-Upgrade and CASTOR but has electron and ion temperatures of the same magnitude. Consequently the ion (and electron) Larmor radius, $\rho_{i}$ on MAST is larger than in the previous two devices, but (unlike TJ-K) the ions remain fully magnetized. This makes MAST an ideal intermediate test of the BPP technique.
\\ \\In section 2 the BPP design for MAST is described. In section 3 plasma potential profile measurements are presented and compared to the standard floating potential measurement from a Langmuir probe. In section 4 the electron temperature calculated from the BPP signals is presented and a comparison with the Thomson scattering diagnostic on MAST is used to determine the success of the BPP measurement. Section 5 presents the radial electric field as measured with the BPP. Section 6 discusses the BPP results before section 7 concludes.

\section{The ball pen probe on MAST}  
\begin{figure}
\includegraphics[width=0.5\textwidth]{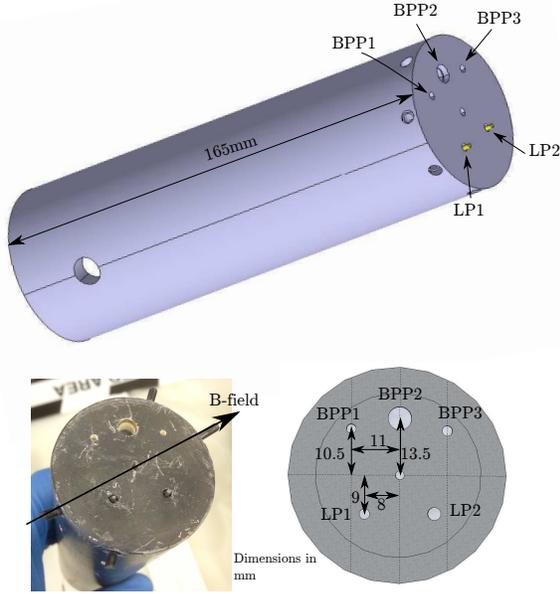}
\centering
\caption{The BPP design for MAST. BPP pins are ball pen probes whilst LP1 and LP2 are standard Langmuir probes. Pins BPP1, BPP2, BPP3 and LP1 are floating pins, whilst LP2 may be externally biased. All unlabeled pins were unused during the experiments commented upon within this paper.}
\label{Fig:BPP_image}
\end{figure}

\begin{figure}[htbp]
\centering
\includegraphics[width=0.7\columnwidth]{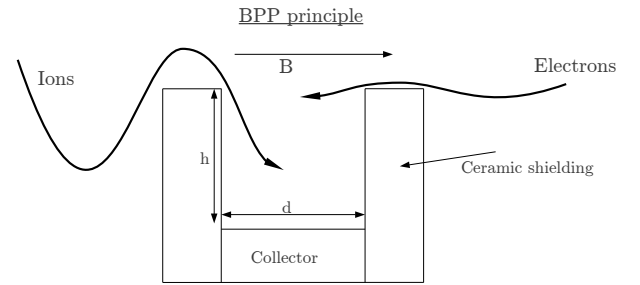}
\caption{Cross section of the ball pen probe principle. The collector is retracted inside a ceramic shielding tube by a distance larger than the ion Larmor radius.}
\label{Fig:BPP_schematic}
\end{figure}

The BPP, shown in figure \ref{Fig:BPP_image}, is not a bespoke design but rather a modification of the Gundestropp probe previously used on MAST \cite{TamainPPCF2010}.
The BPP pins in the probe built for MAST have flat collection surfaces recessed inside a ceramic body, making it similar in design to the BPP used in CASTOR \cite{SchritweisserCJP2006}.  The retraction depths of the BPP pins are fixed below the ion Larmor radius. The essential features defining a BPP are a retracted collector inside a ceramic shielding tube. In the BPP described here the collectors are graphite whilst the shielding tube is boron nitride. Outer surfaces are coated with a thin layer of carbon to prevent excess impurity accumulation in the plasma. Inner surfaces of the shielding tube are bare boron nitride. The dielectric shielding prevents an independent external biasing of the tube from the collector surface. The retraction depths and collector diameters for the three BPP pins on the MAST probe are given in table \ref{Table:Pin_parameters}.
\begin{table}[htbp]
\centering
\begin{tabular}{c c c}
\hline  
Pin & Depth (mm) & Diameter (mm) \\
\hline 
BPP1 & 2 & 1.5\\
BPP2 & 5 & 4 \\
BPP3 & 8 & 1.5
\end{tabular}
\caption{MAST BPP pin retraction depths and diameters. Since the pin retraction and diameter was fixed in each case a suitable range of parameters was chosen for the three pins to allow maximum flexibility.}
\label{Table:Pin_parameters}
\end{table}
Pin LP2 was the only pin attached to an external power supply and could be biased, whilst LP1 as well as the BPP pins were operated in floating mode for the entirety of the experiments presented herein. Of the three BPP pins, BPP1 and BPP3 were found to exhibit suppression of the signal over the entire frequency spectrum as a result of excessively high pin impedences. This could be rectified in future implementations through use of a high impedence (T$\Omega$) electrometer, as used in \cite{AdamekCPP2013}, acting as a voltage buffer. This was noted after the experiments detailed so this modification could not be made at the time. The wider diameter of pin BPP2 significantly reduced the effective pin impedence and the suppression of the low frequency component of the signal was not observed. Suppression of high frequency part of the BPP spectrum was noted in the colder SOL region of the plasma however this did not affect profile measurements.
\\The floating potential of a probe in a plasma, which arises by the constraint of ambipolarity, is given by \cite{Hutchinson}
\begin{equation}
\label{Eqn:Vf}
V_{fl} = \phi - \alpha T_{e}
\end{equation}
where $V_{fl}$ is the floating voltage on the probe, $\phi$ is the plasma potential, $T_{e}$ is the electron temperature in $eV$ and $\alpha = \ln\left(I_{e}/I_{i}\right)$ is the logarithmic ratio of electron to ion currents flowing to the probe. In a magnetized plasma the dominant flux of electrons and ions to an exposed probe, such as a Langmuir probe, occurs along magnetic field lines. This leads to sheath formation as a result of the high electron mobility compared to the ions. Neglecting for now the effect of secondary electron emission, $\alpha_{LP}$ is given in the case of a Langmuir probe by \cite{Stangeby}
\begin{equation}
\label{Eqn:alpha}
  \alpha = \ln\left(\frac{I_{e}}{I_{i}}\right) = -\frac{1}{2}\ln\left(2\pi\frac{m_{e}}{m_{i}}\left(1+\frac{T_{i}}{T_{e}}\right)\right)
\end{equation}
The goal of the BPP is to reduce $\alpha$ to zero by equalizing the ratio of electron to ion currents at the collector. This is achieved by shielding the collector from fluxes directed along the magnetic field line such that cross-field transport is required for the net collection of ions and electrons. This principle is illustrated in figure \ref{Fig:BPP_schematic}.
The results of 2D particle-in-cell simulations \cite{KommCPP2010,KommPPCF2011} of a BPP suggest that equation \ref{Eqn:Vf} holds as the BPP is retracted inside the probe surface, and significant experimental evidence supports the use of equation \ref{Eqn:Vf} in a BPP. For the purposes of this paper equation \ref{Eqn:Vf} will be assumed to apply in the absense of a first principles model for the BPP. Since no current model exists to predict $\alpha_{BPP}$ this must be treated as a free parameter in this study. In previous studies $\alpha_{BPP}$ has been measured directly\cite{AdamekCJP2004,AdamekCJP2005,SchritweisserCJP2006} by applying an external voltage sweep to the BPP, however this is not possible in the present implementation. Instead $\alpha_{BPP}$ will be constrained by a detailed cross-diagnostic comparison with the Thomson scattering system on MAST\cite{ScannelRSI2006}. The BPP has been compared favourably with potential measurements from an emissive probe\cite{AdamekCJP2005} and a self-emitting Langmuir probe\cite{AdamekCPP2014}.
\\The results in this paper are drawn from three connected double-null (CDN) Ohmic L-mode plasmas on MAST. These are shot numbers 28819, 28830 and 28834. The relevant details of these shots are given in table \ref{Table:Shotparameters}.
\begin{table}[htp]
\begin{tabular}{c c c c c}
\hline
Shot & $I_{p} (kA)$ & $B_{T} (T)$ & $n_{e,LCFS} (10^{19}m^{-3})$ & $\overline{n_{e}}(10^{18}m^{-2})$ \\
\hline
28819 & 400 & 0.585 & 0.8 & 120 \\ 
28830 & 600 & 0.585 & 0.8 & 110 \\ 
28834 & 600 & 0.585 & 0.3  & 70 
\end{tabular}
\centering
\caption{Plasma parameters for MAST shots 28819, 28830 and 28834. $I_{p}$ is the plasma current, $B_{T}$ is the toroidal field on axis, $n_{e,edge}$ is the plasma density at a major radius of $1.38m$ which corresponds to the end of the probe reciprocation and $ \overline{n_{e}}$ is the line integrated plasma density.}
\label{Table:Shotparameters}
\end{table}
The three shots represent both a two-point plasma current and a two-point electron density scan Ohmic L-mode. The BPP was attached to the end of the MAST reciprocating probe (RP) system \cite{YangJNM2003} which has a slow motorized drive for initial placement of the probe, and a fast pneumatic drive for reciprocations into the plasma. The maximum velocity attainable by the probe head is $1m/s$. In the CDN configuration the RP is located at the outboard midplane and each reciprocation occurs over a distance of $8cm$ in major radius. The motion of the plasma edge is accounted for using an EFIT\cite{EFIT} reconstruction and profiles will be presented in terms of $R - R_{LCFS}$, where $R$ is the major radius and $R_{LCFS}$ is the radial location of the last closed flux surface. This is subject to a systematic error of $\leq 2cm$ however this error is included in all results and is accounted for when making cross-diagnostic comparisons, so has no significant impact on conclusions in this paper.
\\To construct radial profiles of measured quantities raw signals are split into 50 temporal bins with a width of $2.4$ms. On this time-scale the RP is stationary allowing an average of the data to be taken within each bin. Each bin contains $\approx 1200$ data points which reduces the error on the mean to an insignificant level. Any systematic offset errors have been accounted for by calibration against a control case where no plasma is present.  
\section{BPP Potential Measurements}
\begin{figure}[htbp]
\centering
\includegraphics[width=0.6\columnwidth]{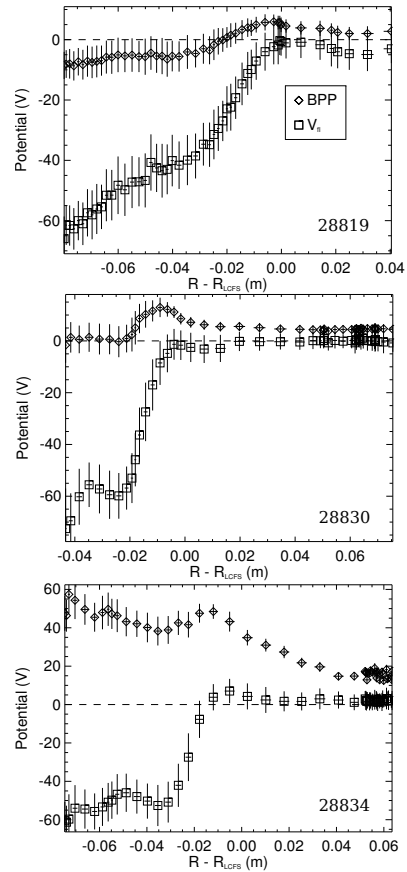}
  \caption{Potential profiles measured by the BPP (diamonds) and the LP (squares). Error bars show the variance within each bin which is dominated by random fluctuations. The error on the mean is $\approx 35$ times smaller than the error bars shown.}
\label{Fig:Potential}
\end{figure}

Figure \ref{Fig:Potential} shows the radial variation of the potential measured by the BPP (diamonds) and by the LP (squares) during reciprocation into the plasma in shots 28819, 28830 and 28834. The error bars indicate the variance of the data within the bin, not the error on the mean. The random error on the mean is approximately 35 times smaller than the error bars shown.
 The first and most striking observation to be made from figure \ref{Fig:Potential} is that the BPP signal cannot simply be a suppressed version of the LP signal. In all three shots the BPP shows regions of opposite polarity to the floating potential. This fits with the expectation from equation (\ref{Eqn:Vf}) that the floating potential should always be more negative than the BPP signal. Secondly the radial profile of the BPP signal differs with respect to that of the floating potential.  The BPP and $V_{fl}$ profiles diverge as the probe penetrated the LCFS, in agreement with expression \ref{Eqn:Vf} where an increasing electron temperature should increase the deviation between $V_{BPP}$ and $V_{fl}$. 
\\The measured BPP signal demonstrates the qualitative features suggested by expression \ref{Eqn:Vf}. To test the validity of the BPP measurement quantitatively a comparison will be made with the electron temperature measurement from the Thomson scattering diagnostic \cite{ScannelRSI2006}.
\section{Electron Temperature Measurements}
\begin{figure}[htbp]
  \centering
\includegraphics[width=0.6\columnwidth]{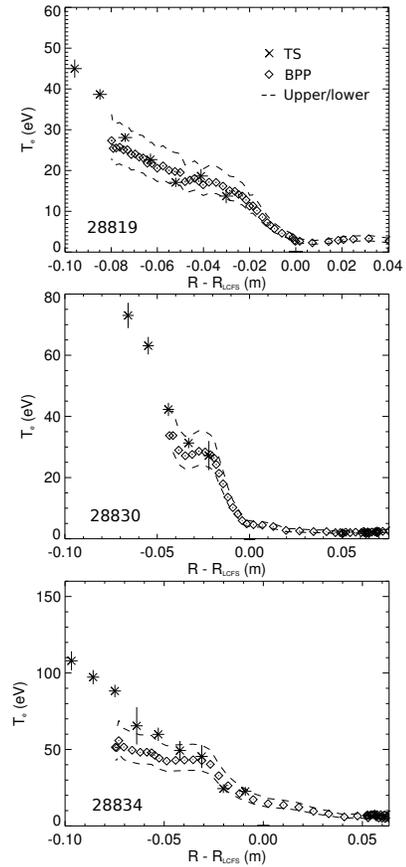}
  \caption{Electron temperature measured by the BPP (diamonds) with upper and lower bounds (lines) compared to the TS measurement (stars). Upper and lower bounds are obtained by taking the extreme values from expression \ref{Eqn:alphauncert}.}
\label{Fig:BPPTeLCFS}
\end{figure}

\begin{figure}
\centering
\includegraphics[width=\columnwidth]{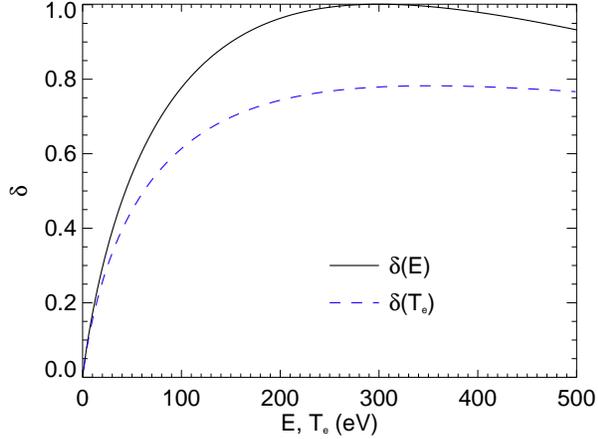}
\caption{Secondary electron yield as a function of energy, $\delta\left(E\right)$ calculated from expression \ref{Eqn:de}, and as a function of electron temperature, $\bar{\delta}\left(T_{e}\right)$ calculated by integration over a Maxwellian.}
\label{Fig:delta}
\end{figure}

\begin{figure}[htbp]
\includegraphics[width=0.6\columnwidth]{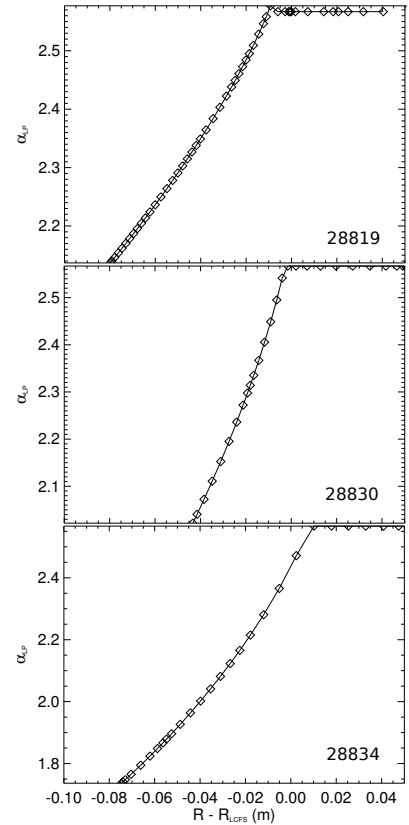}
\caption{Profiles of $\alpha_{LP}$ including the effects of secondary electron emission.}
\label{Fig:alpha_LP}
\end{figure}

\begin{figure}[htbp]
\includegraphics[width=0.62\columnwidth]{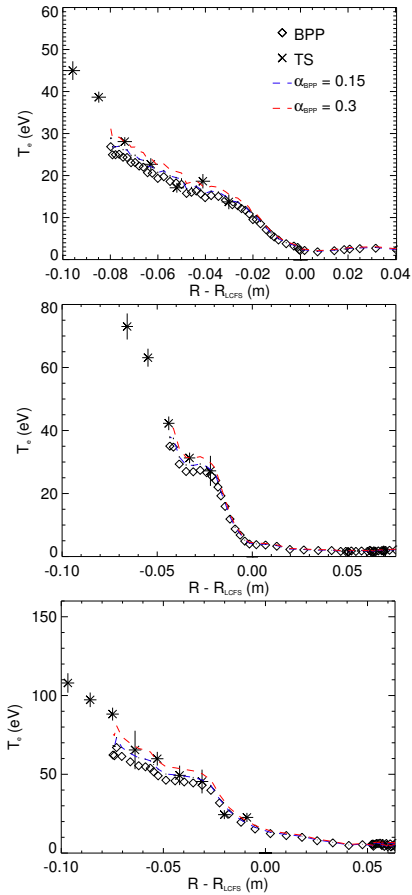}
\caption{BPP measurements of $T_{e}$, accounting for secondary electron emission, compared to TS data (crosses). The BPP data shows the case of $\alpha_{BPP} = 0$ (diamonds), $\alpha_{BPP} = 0.15$ (bue line) and $\alpha_{BPP} = 0.3$ (red line).}
\label{Fig:BPP_Te_2}
\end{figure}

Taking equation \ref{Eqn:Vf} and subtracting the floating potential from the BPP from the floating potential of an LP gives an estimate of the electron temperature, $T_{e}$ of 
\begin{equation}
\label{Eqn:TeBPP}
T_{e} \approx \frac{V_{BPP} - V_{fl}}{\alpha_{LP} - \alpha_{BPP}}
\end{equation}
The uncertainty on both $\alpha_{BPP}$ and $\alpha_{LP}$ leads to a range of values of the denominator in equation \ref{Eqn:TeBPP}. Uncertainty in $\alpha_{LP}$, given in equation \ref{Eqn:alpha}, arrises from uncertainty in $T_{i}/T_{e}$. Previous measurements on MAST have shown that $1 \leq T_{i}/T_{e} \leq 2.5$ is reasonable for L-mode conditions\cite{ElmorePPCF2012,ElmoreJNM2013,AllanJNM2013}.This then gives $\alpha_{LP} = 2.7\pm 0.1$. Since $\alpha_{BPP}$ cannot be measured directly in the system described in this paper the empirical value of $\alpha_{BPP} = 0.6\pm 0.3$\cite{AdamekCJP2004,AdamekJNM2009,HoracekNF2010,AdamekEPS2014} found across a number of experiments is adopted. Recently this approach has produced electron temperature profiles on COMPASS and AUG in close agreement with thomson scattering measurements \cite{AdamekEPS2014}. Taking these values of $\alpha_{LP}$ and $\alpha_{BPP}$ leads to a range of values for the denominator of equation \ref{Eqn:TeBPP} of
\begin{equation}
\label{Eqn:alphauncert}
\alpha_{LP} - \alpha_{BPP} = 2.1 \pm 0.4
\end{equation}
which is used here to give a central, maximal and minimal temperature measured from the BPP. MAST has a well resolved (spatially and temporally) Thomson scattering (TS) diagnostic \cite{ScannelRSI2006} which is used as a comparison with the electron temperature derived from the BPP data. The TS data is sampled in a phase of the plasma shot where the edge is approximately stationary. The data is binned in radius and averaged with error bars representing the standard error on the mean. Error on the TS data is a combination of random error associated with the measurement technique and the variance of fluctuations within the binning window. Data points with a random error greater than half the signal were rejected (in practice this was not a significant portion of the data set). The comparison between the BPP and the TS diagnostic is shown in figure \ref{Fig:BPPTeLCFS}.
\\The TS data lies within the range of uncertainty of the BPP data in each case. Whilst this is encouraging the uncertainty on $\alpha_{LP} - \alpha_{BPP}$ is large and results primarily from uncertainty on $\alpha_{BPP}$. Furthermore the BPP measurement appears to be systematically lower than the TS measurement, with the disparity between the two measurements increasing at higher electron temperatures. By comparing the BPP measurement with the TS measurement $\alpha_{BPP}$ can be constrained independently of the empirical value derived elsewhere. In order to accurately constrain $\alpha_{BPP}$ however, $\alpha_{LP}$ must be specified as accurately as possible. In particular the effects of secondary electron emission have thus far been neglected. This is common practice in the case of flush-mounted probes where the magnetic field line incidence onto the probe is shallow and any emitted electrons are recaptured within a Larmor orbit. In the present case however the magnetic field has a near normal incidence on the probe surface and emitted electrons are accelerated away along the magnetic field line, as a result secondary electron emission may be an important effect. Including the effects of secondary electron emission, $\alpha_{LP}$ is given by \cite{Stangeby}
\begin{equation}
\label{Eqn:alpha_de}
\alpha_{LP} = -\frac{1}{2}\ln\left(2\pi\frac{m_{e}}{m_{i}}\left(1+\frac{T_{i}}{T_{e}}\right)\left(1-\delta_{e}\right)^{-2}\right)
\end{equation}
where $\delta_{e}$ is the ratio of secondary electrons emitted per primary incident electron. $\delta_{e}$ can be calculated for a graphite probe tip from the semi-empirical formula \cite{Stangeby,ThomasNF92}
\begin{equation}
\label{Eqn:de}
\delta_{e} = 2.72^{2}\frac{E}{300}\exp\left(-2\left(\frac{E}{300}\right)^{1/2}\right)
\end{equation}
where $E$ is the incident electron energy. From this expression the secondary electron emission as a function of electron temperature, $\bar{\delta}\left(T_{e}\right)$ can be obtained by integration over a Maxwellian
\begin{equation}
\bar{\delta}\left(T_{e}\right) = \frac{\int_{0}^{\infty}\delta\left(E\right)\exp\left[-E/T_{e}\right]dE}{\int_{0}^{\infty}\exp\left[-E/T_{e}\right]dE}
\end{equation}
where $E$ is in eV. Note that $\delta$ has been assumed here to be constant in pitch angle. Figure \ref{Fig:delta} shows $\delta\left(E\right)$ and $\bar{\delta}\left(T_{e}\right)$.
 Experimentally the secondary electron emission has been shown to increase as the angle of incidence reduces\cite{WoodsJPD87}, however no parameterization of this effect exists. As such expression \ref{Eqn:de}, which relates formally to normal incidence electrons, must be used. It should be noted though that this then gives an underestimate of the effect of secondary electron emission. 
\\With the secondary electron yield parameterized as a function of $T_{e}$, $\alpha_{LP}$ can be calculated. This is achieved by taking a linear regression of the TS data, shown in figure \ref{Fig:BPPTeLCFS}, to the point where $T_{e} < 1$eV, beyond which the profile is set at $T_{e} = 1$eV. This captures the main aspects of the temperature variation during the probe reciprocation and more detailed fits to the data were found to be unnecessary. This approximation to $T_{e}$ is used as input to equation \ref{Eqn:alpha_de} and a radial profile of $\alpha_{LP}$ is produced. Figure \ref{Fig:alpha_LP} shows the profile of $\alpha_{LP}$ obtained with this technique in each shot. In these calculations $T_{i} = 2T_{e}$ was assumed, which is motivated by previous measurements on MAST \cite{ElmorePPCF2012,ElmoreJNM2013,AllanJNM2013}.
Figure \ref{Fig:alpha_LP} shows that the secondary electron emission can have a major impact on the calculation of $\alpha_{LP}$. It also causes $\alpha_{LP}$ to vary radially, which can affect the profile shape of the temperature measurement. It should be noted that a radial variation of $\alpha_{BPP}$ cannot be ruled out without a fuller understanding of the BPP collection mechanism, however direct measurements of $\alpha_{BPP}$ have yet to reveal such a radial variation \cite{AdamekCJP2004}. The values of $\alpha_{LP}$ can now be used to re-calculate the BPP measurement of $T_{e}$. This is shown in figure \ref{Fig:BPP_Te_2}.
\\By accounting for secondary electron emission the BPP measurement of $T_{e}$ agrees well with the TS data. The majority of the TS data, particularly in the hotter region inside the separatrix, lies between the BPP measurements obtained with $\alpha_{BPP} = 0$ (diamonds in figure \ref{Fig:BPP_Te_2}) and $\alpha_{BPP} = 0.3$ (red, broken line in figure \ref{Fig:BPP_Te_2}). Outliers to this trend occur in the lowest temperature region, near to the separatrix and SOL. In these regions the TS data hits a floor measurement of 5eV which can artificially raise the TS average. Furthermore the random error on the TS data is much larger in these regions. The more reliable region of comparison occurs towards the hotter end of the profile.  Assuming that the true value of $\alpha_{BPP}$ remains constant between all three shots, these results suggest that $\alpha_{BPP}$ for the BPP on MAST can be constrained by $ 0 \leq \alpha_{BPP} \leq 0.3$. This is at the lower end of the values previously recorded in literature. This may be due to the lack of a conical collector, with the MAST collector being entirely flat. The conical shape allows for a portion of electron and ion currents to intersect the collector with a normal incidence along the field line. By removing the conical shape the transport to the collector must occur perpendicularly to the magnetic field and is likely to be closer to ambipolar than the conical design. This may result in a reduced $\alpha_{BPP}$ in the case of a flat collector, as observed here. As indicated earlier, the lack of accounting for pitch angle in the calculation of the secondary electron emission suggests that these results are underestimates of the true BPP temperature. Accounting for pitch angle can provide an enhancement to the secondary electron emission by up to $\sim 40$\%\cite{WoodsJPD87}. Applying an artificial enhancement of this magnitude to the secondary electron emission was observed to bring the BPP and TS measurements into close agreement with $\alpha_{BPP} \approx 0$. 
\section{Measurements of the radial electric field}
\begin{figure}[htbp]
\centering
\includegraphics[width=0.6\columnwidth]{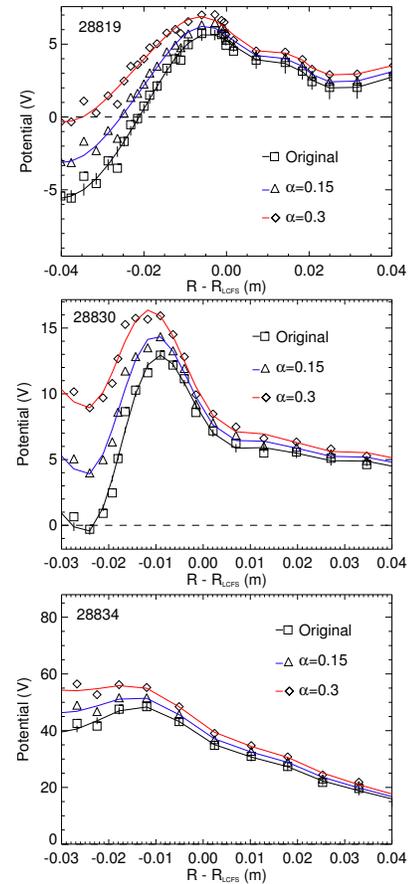}
\caption{Corrected BPP measurements of the plasma potential taking $\alpha_{BPP} = 0$ (diamonds), $\alpha_{BPP} = 0.15$ (blue) and $\alpha_{BPP} = 0.3$ (red). The uncertainty on the potential measurement grows as the probe moves deeper into the plasma, where the electron temperature increases and the consequential correction to the BPP potential grows. Solid lines show a 17$^{th}$ degree polynomial fit, used to obtain continuous profiles with a view to taking the derivative.}
\label{Fig:BPP_correct}
\end{figure}

\begin{figure}[htbp]
\centering
\includegraphics[width=0.6\columnwidth]{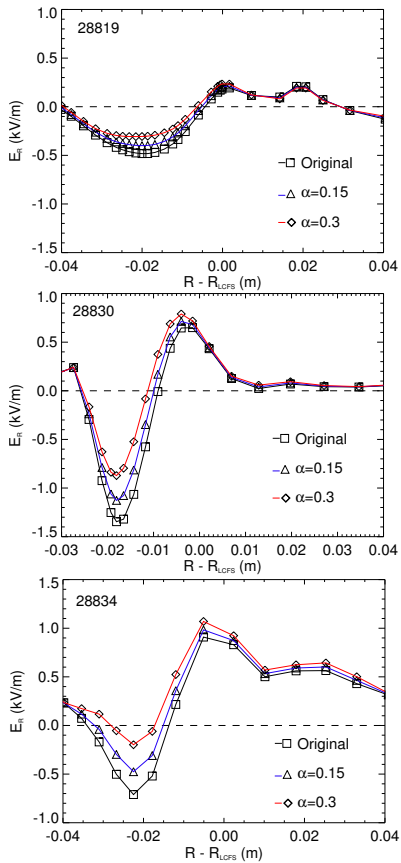}
  \caption{Estimates of the radial electric field obtained by fitting a high order polynomial to the data presented in figure \ref{Fig:BPP_correct} and taking $E_{r}\approx -\partial_{r}V_{BPP}$ for the correction factors $\alpha_{BPP} = 0$ (black), $\alpha_{BPP} = 0.15$ (blue) and $\alpha_{BPP} = 0.3$ (red). The greatest uncertainty in the $E_{R}$ occurs in the negative well on the inside of the radial shear layer.}
\label{Fig:Efield}
\end{figure}

\begin{figure}[htbp]
\centering
\includegraphics[width=0.6\columnwidth]{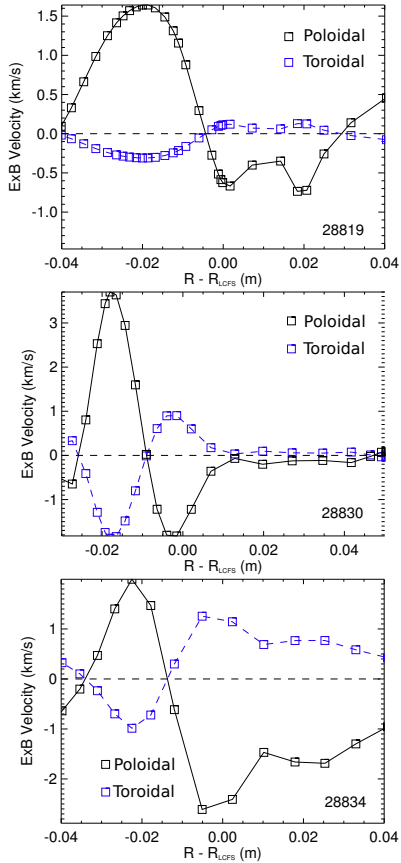}
  \caption{Toroidal (solid) and poloidal (broken) velocities in the shear layer inferred from the $E_{r}$ estimates in figure 8 and using $v_{pol}\approx E_{r}/B_{tor}$ and $v_{tor} \approx E_{r}/B_{pol}$. $B_{tor}$ and $B_{pol}$ are taken from and EFIT reconstruction.}
\label{Fig:Velocities}
\end{figure}

By cross-diagnostic comparison the BPP has been shown to make a direct measurement of a quantity which can be (at least approximately) identified with the plasma potential. The plasma potential in it self is an interesting measurement, however a more practically useful quantity is the radial electric field. The radial shear layer, a region of strong velocity shear near to the LCFS, is likely to play a key role in the anomalous transport of particles and energy into the SOL\cite{XuNF2009}. To leading order the velocity in the shear layer results from $\textbf{E}\times\textbf{B}$ motion so is predominantly characterized the radial electric field it self. This provides strong motivation for accurate measurements of the radial electric field. 
\\ \\The BPP signal can be corrected to measure the true plasma potential by the relation
\begin{equation}
\label{Eqn:Phicorrected}
\phi = V_{BPP} + \alpha_{BPP}T_{e} = \left(1 + \gamma\right)V_{BPP} - \gamma V_{fl}
\end{equation}
where  $\gamma = \alpha_{BPP}/\left(\alpha_{LP} - \alpha_{BPP}\right)$. Clearly if $\alpha_{BPP} = 0$ the BPP measurement of the plasma potential is exact, however given the uncertainty in the experimental comparison with the TS system, it is not possible to eliminate the possibility of finite $\alpha_{BPP}$. As such it is important to assess how this uncertainty may affect measurements of the radial electric field. The BPP signal can be corrected by using the BPP calculation of $T_{e}$ in equation \ref{Eqn:Phicorrected}. Figure \ref{Fig:BPP_correct} shows the corrected BPP measurement of the plasma potential, obtained by taking $\alpha_{BPP} = 0, 0.15$ and $0.3$ respectively.
The effect of varying $\alpha_{BPP}$ on the plasma potential increases as the electron temperature grows and the corresponding correction to $\phi$ grows. The difference between the measured potential profiles can be significant, which is particularly true in shot 28830. To see how this uncertainty affects the radial electric field measurements, the radial electric field is taken as the derivative of the BPP profiles in figure \ref{Fig:BPP_correct} with respect to $\hat{R} = R - R_{LCFS}$ (to negate the effect of plasma edge motion) such that $E_{R} = -\nabla\phi = -\partial\phi/\partial \hat{R}$. To take the derivative the BPP profile is fitted to a $17^{th}$ degree polynomial (fits shown as solid lines in figure \ref{Fig:BPP_correct}) and the derivative is subsequently taken. Since no functional form of the radial electric field at the LCFS is present in the literature a fitting function was chosen that provided a smooth profile, whilst capturing all of the important profile features. The radial electric field measurements, constructed in this manner, for shots 28819, 28830 and 28834 are shown in figure \ref{Fig:Efield}.
The structure of the radial electric field measured on MAST agrees qualitatively with BPP and Doppler Reflectometry measurements on ASDEX-Upgrade\cite{MullerNF2011}. In each shot a shear-layer can clearly be identified where the radial electric field changes sign.  The magnitude of the electric field in the shear layer is on the order of $\sim1$kV/m which agrees with previous studies on MAST \cite{MeyerJNM2008,TempleThesis}. The uncertainty induced by the variation in $\alpha_{BPP}$ in the measurement of $E_{R}$ is significantly less pronounced than in the plasma potential measurement. The uncertainty is most significant in the negative electric field well that occurs on the inside of the radial shear layer, where it can reach a level $\sim \pm 0.4$kV/m, however the uncertainty is reduced below this level over much of the profile. As a consequence the BPP can be considered to make a well resolved estimate of $E_{R}$. Indeed, since the qualitative features of the radial electric field are independant of $\alpha_{BPP}$, to a good approximation the raw BPP data can be used to infer the radial electric field if $\alpha_{BPP}$ is unknown. The results in figure \ref{Fig:Efield} show a systematic increase in the magnitude of the radial electric field between shot 28819, and shots 28830 and 28834. The principle difference between these two shots is the plasma current (see table \ref{Table:Shotparameters}) hinting that the radial electric field increases in magnitude (and consequently shear) with plasma current. The relative similarity between the measurements in shots 28830 and 28834 in both magnitude and shear show that increased density seems to play a lesser role in the formation of the radial electric field. The BPP offers the opportunity for further study of the scaling behaviour of the radial electric field and, of particular interest, the radial electric field shear and may therefore be a valuable tool in the effort to understand plasma transport in the tokamak periphery. 
\\ \\Using the radial profile of the toroidal and poloidal magnetic field taken from an EFIT reconstruction, a poloidal and toroidal $\textbf{E}\times\textbf{B}$ velocity estimate can be inferred respectively. By separating the poloidal and toroidal components of $\textbf{E}\times\textbf{B}/B^{2}$ the velocity components can be given by
\begin{equation}
v_{\theta} = -\frac{B_{\zeta}E_{R}}{B^{2}} \approx \frac{B_{\zeta}}{B^{2}}\frac{\partial V_{BPP}}{\partial R}
\end{equation}
\begin{equation}
v_{\zeta} = \frac{B_{\theta}E_{R}}{B^{2}} \approx -\frac{B_{\theta}}{B^{2}}\frac{\partial V_{BPP}}{\partial R}
\end{equation}
 where $B_{\theta}$ is the poloidal magnetic field strength, $B_{\zeta}$ is the toroidal magnetic field strength and $B$ is the total magnetic field strength. The calculated $v_{\theta}$ and $v_{\zeta}$ are shown in figure \ref{Fig:Velocities}. The velocities are calculated using the $E_{R}$ profiles with $\alpha_{BPP} = 0$ and represent velocity profile estimates containing the maximum possible shear from the BPP data. Radial profiles of $B_{\zeta}$, $B_{\theta}$ and $B$ are used in these calculations, however minimal variation is observed across the range of radii used. 
The plasma is observed to rotate poloidally and toroidal with a magnitude on the order of $\sim 1$km/s. The comparative increase in the poloidal component over the toroidal component is a result of the weaker poloidal magnetic field. The shear in the radial electric field leads to a velocity shear layer which can be clearly identified in each shot once again about the point where the velocity changes sign. In the future it is hoped that measurements such as these can be compared to other diagnostic systems on MAST, DBS (Doppler back-scattering) for example, to both increase the accuracy of velocity measurement on MAST and help to further constrain the uncertainty in the BPP measurements.

\section{Conclusions}
The ball pen probe technique has been successfully used on MAST to make profile measurements of potential, electron temperature and radial electric field. It has been shown to make a measurement which lies much closer to the true plasma potential than the floating potential and is demonstrated to behave differently to the floating potential both in terms of polarity and profile shape. By combining the BPP measurement with the standard measurement of floating potential from a Langmuir probe the electron temperature profile has been derived and compared favourably to a complementary measurement made by the MAST Thomson scattering system. This constrains the parameter $\alpha_{BPP}$ to $0\leq\alpha_{BPP}\leq 0.3$. In making the diagnostic comparison between the BPP and the TS system, accounting for secondary electron emission was shown to be important in constraining $\alpha_{BPP}$. More generally this suggests that the effect of secondary electron emission should be accounted for when interpretting probe data where the field line incidence to the probe is close to normal. By fitting a polynomial function to the BPP potential profiles and taking the derivative with respect to distance from the LCFS the radial electric field is measured for Ohmic L-mode MAST plasmas. It is shown to exist on the order of $\sim 1 kV/m$ and appears to increase with increasing plasma current. Uncertainty in the measurement is relatively low but reaches its highest point in the negative well of the electric field, inside the radial shear layer. From these estimates a toroidal and poloidal rotation velocity can be inferred by taking appropriate values of the poloidal and toroidal magnetic field respectively.
\\This paper highlights the capability of the ball pen probe to make several useful and important measurements in the plasma edge. Furthermore it has a robust design capable of operation well inside the LCFS. The specific design used for MAST was a straightforward modification of an existing probe-head showing the ease with which a ball pen probe can be implemented. A drawback of the technique is an imprecise understanding of its collection mechanism, however it is hoped that the empirical success of the ball pen probe motivates continued use of this technique in the study of boundary plasma phenomena. 
 
\section{Acknowledgements}
This work was part-funded by the RCUK Energy Programme under grant EP/I501045, by project GA CR P205/12/2327 of the Grant Agency of Czech Republic, by project MSMT LM2011021 and by the European Communities. To obtain further information on the data and models underlying this paper please contact PublicationsManager@ccfe.ac.uk. The views and opinions expressed herein do not necessarily reflect those of the European Commission.

\section{References}
\bibliographystyle{prsty}
\bibliography{Paper-ref}

\end{document}